\documentclass[twocolumn,tightenlines]{revtex4}   
\usepackage{amsfonts,amssymb,amsmath,amscd}
\usepackage{dsfont}
\usepackage{framed}

\usepackage{mathrsfs}
\usepackage{wasysym}
\usepackage{stmaryrd}
\usepackage{rotating}
\usepackage{txfonts}
\usepackage{pifont}

\begin{document}

\title{A solution for  the  paradox of the double-slit experiment}

\author{Gerrit Coddens}

\affiliation{                    
 Laboratoire des Solides Irradi\'es, 
Universit\'e Paris-Saclay,\\ F-91128-Palaiseau CEDEX, FRANCE
}

\begin{abstract}
We argue that the double-slit experiment can be understood much better by considering it as an experiment whereby
one uses electrons to study the set-up rather than an experiment whereby we use a set-up to study the behaviour of electrons.
We also show how  the concept of undecidability  can be used in an intuitive way to make sense of the double-slit experiment
and the quantum rules for calculating coherent and incoherent probabilities. We meet here a situation where the electrons always behave in a fully
deterministic way (following Einstein's conception of reality), while the detailed design of the set-up may render the question about the way they move 
through the set-up  experimentally undecidable (which follows more Bohr's conception of reality).
We show that the expression $\psi_{1} + \psi_{2}$ for the wave function of the double-slit experiment 
 is numerically correct, but logically flawed. It has to be replaced in the interference region
by the logically correct expression $\psi'_{1} + \psi'_{2}$, which has the same numerical value as 
$\psi_{1} + \psi_{2}$, such that $\psi'_{1} + \psi'_{2} = \psi_{1} + \psi_{2}$,
but with $\psi'_{1} =  {\psi_{1} +\psi_{2}\over{\sqrt{2}}} \,e^{\imath {\pi\over{4}}  } \neq \psi_{1}$ and 
$\psi'_{2} =    {\psi_{1} +\psi_{2}\over{\sqrt{2}}}\,e^{-\imath {\pi\over{4}}}\neq \psi_{2}$.
Here $\psi'_{1}$ and $\psi'_{2}$ are the correct
contributions from the slits to the total wave function  $\psi'_{1} + \psi'_{2}$.  
We have then $p = |\psi'_{1} + \psi'_{2}|^{2} = |\psi'_{1}|^{2} + |\psi'_{2}|^{2} = p_{1}+p_{2} $ such that the paradox 
that quantum mechanics (QM) would not follow the traditional rules of probability calculus disappears.
 The paradox is rooted in the wrong intuition that  $\psi_{1}$ and $\psi_{2}$ would be the true physical contributions to 
$\psi'_{1} + \psi'_{2} =\psi_{1} + \psi_{2}$ like in the case of waves in a water tank. 
The solution proposed here is not {\em ad hoc} but based on an extensive analysis of the geometrical meaning of spinors within  group representation  theory
and its application to QM. Working further on the argument one  can even show that an interference pattern is the only way to satisfy simultaneously
 two conditions: The  condition obeying binary logic  (in the spirit of Einstein) that the electron  has only two mutually exclusive options to get to the detector (viz. going through slit S$_{1}$ or going through slit S$_{2}$)  and the  condition obeying ternary logic (in the spirit of Bohr)
 that the question which one of these two options the electron has taken is experimentally undecidable.
\end{abstract}

\maketitle

\section{Introduction} \label{intro}

The double-slit experiment has been qualified by Feynman \cite{Feynman1} as the only mystery of QM.
Its mystery resides in an apparent paradox between the QM result and what we expect  on the basis of our intuition.
What we want to explain in this article is that this apparent paradox is a probability paradox.
By this we mean that the paradox does not  reside in some special property of the electron that could act
both as a particle and a wave, but in the fact
 that we use two different definitions of probability in the intuitive approach and in the calculations.
 It is the difference between these two definitions which leads to the paradox, because the two definitions are just incompatible.
 In our discussion we will very heavily rely on the presentations by Feynman, even though further strange aspects have been 
 pointed out by other authors later on, e.g. in the discussion of the delayed-choice experiment by Wheeler \cite{Wheeler}
 and of the quantum eraser experiment \cite{eraser}, which can also be understood based on our discussion.
 The afore-lying paper is an introduction to a much longer and detailed paper \cite{Coddens-HAL}, which  
 itself relies heavily on a complete monograph \cite{Coddens}. 

 \section{Feynman's essentials} \label{Feyn}
 
 Feynman illustrates the paradox by comparing tennis balls and electrons.  Tennis balls comply  with classical intuition,
 while electrons behave according to the rules of QM. There is however,  a small oversimplification in Feynman's discussion. 
He glosses over a detail, undoubtedly for didactical reasons. When the electron behaves quantum mechanically and only one slit is open,
 the experiment will give rise to diffraction fringes, which can also not be understood in terms of a classical description in terms of tennis balls.
But the hardest part of the mystery is  that in the quantum mechanical regime we get a diffraction pattern when only one slit is open, while we get an interference pattern
 when both slits are open. This means that  the single-slit probabilities even do not add up to an interference pattern when we allow for the quantum
 nature of  the electron in a single-slit experiment.
 We will therefore  compare most of the time the two quantum mechanical situations  rather than electrons and tennis balls.

What Feynman describes very accurately is how quantum behaviour  corresponds to the idea that the electron does not leave any
trace behind in the set-up of its interactions with it, that would permit to reconstruct its history. (We exclude  here from our concept of a set-up the detectors that register the electrons at the very end
of their history). We cannot tell with what part of the set-up the electron has interacted, because the interaction has been coherent.
This corresponds to ``wave behaviour''.
At the very same energy, a particle can also  interact incoherently with the set-up and this will then result in classical ``particle behaviour''.
The difference is that when the particle has interacted incoherently we do have the possibility to figure out its path trough the device, because
the electron has left behind indications of its interactions with the measuring device  within the device.

A nice example of this difference between coherent and incoherent interactions occurs in neutron scattering. 
In its interaction with the device, the neutron can flip its spin. The conservation of angular momentum
implies then that there must be a concomitant change of the spin of a nucleus within an atom of the device. At least in principle the
change of the spin  of this nucleus could be detected by comparing the situations before and after the passage of the neutron, such that the history of the neutron
could be reconstructed. 
Such an interaction with spin flip corresponds to incoherent neutron scattering.
But the neutron can also interact with the atom without flipping its spin. There will be then no trace of the passage of the neutron in the form of a change of spin
of a nucleus, and
we will never be able to find out the history of the particle from a {\em post facto} inspection of the measuring device.
An interaction without spin flip corresponds to coherent scattering.
Note that this discussion only addresses  the coherence of the spin interaction.
There are other types of interaction possible and in order to have a globally coherent process none of these interactions
must leave a mark of the passage of the neutron in the system that could permit us to reconstruct its history. An example of an alternative
distinction between coherent and incoherent scattering occurs in the discussion of the recoil of the atoms of the device. A crystal lattice can recoil as a whole (coherent scattering). Alternatively, the recoil can just affect a single atom (incoherent scattering).

 In incoherent scattering the electron behaves like a tennis ball. The hardest part of the mystery of the double-slit experiment is thus the paradox which occurs when we compare coherent scattering in the single-slit and in the double-slit experiment.
Feynman resumed this mystery by asking: How can the particle know if the other slit is open or otherwise?
In fact, as its interactions  must be local the electron should not be able to ``sense'' if the other slit is open (see below).

\section{Caveats} \label{cave}

Let us now leave our intuition for what it is and turn to QM. To simplify the formulation, we will in general use the term probability for what are in reality probability densities.
In a purely QM approach we could make the calculations for the three configurations of the experimental set-up. We could solve the wave equations
for the single-slit and  double-slit experiments:

\begin{equation} \label{slit1}
\begin{tabular}{lll}
$ -{\hbar^{2}\over{2m}} \Delta \psi_{1} + V_{1}({\mathbf{r}}) \,\psi_{1} = - {\hbar\over{\imath}} {\partial\over{\partial t}} \psi_{1},$ &   S$_{1}$ open, &  S$_{2}$ closed,\\
$ -{\hbar^{2}\over{2m}} \Delta \psi_{2} + V_{2}({\mathbf{r}})\,\psi_{2} = - {\hbar\over{\imath}} {\partial\over{\partial t}} \psi_{2},$ & S$_{1}$  closed, &  S$_{2}$ open,\\
$  -{\hbar^{2}\over{2m}} \Delta \psi_{3} + V_{3}({\mathbf{r}})\, \psi_{3} = - {\hbar\over{\imath}} {\partial\over{\partial t}} \psi_{3},$ &   S$_{1}$ open, &  S$_{2}$ open.\\
 \end{tabular}
 \end{equation}
  
\noindent Here S$_{j}$ refer to the slits. Within this theoretical framework we would still not obtain the result $|\psi_{3}|^{2}$ for the double-slit
experiment by adding the probabilities $|\psi_{1}|^{2}$ and $|\psi_{2}|^{2}$ obtained from the solutions of the wave equations for the single-slit experiments.
The fact that $|\psi_{3}|^{2} \neq  |\psi_{1}|^{2}+ |\psi_{2}|^{2}$ is at variance with our intuition about the rules of probability calculus
in a way that seems to defy all our logic, because we expect the electron to have only two mutually exclusive options. It must travel through
S$_{1}$ or through S$_{2}$.
Textbooks tell us that we should not add up probabilities but probability amplitudes,  $|\psi_{3}|^{2} =  |\psi_{1} + \psi_{2}|^{2}$. They describe this as
the ``superposition principle''. They define  wave functions $\psi  = \sum_{j} c_{j} \chi_{j}$, and corresponding
probabilities $p = |\psi|^{2} = | \sum_{j} c_{j} \chi_{j}|^{2}$, whereby one must combine probability amplitudes rather than probabilities
in a linear way  (coherent summing).
They compare this to the addition of the amplitudes of waves like we can observe
 in a water tank, as also discussed by Feynman.
 
  It must be pointed out that adding wave functions is certainly algebraically feasible, but {\em a priori} incompatible
 with their geometrical meaning. Wave functions are spinors or simplified versions of them and spinors in representation theory have a well-defined geometrical meaning, physicists are not aware of. We can draw an analogy between the situation in QM  and what happens in algebraic geometry, where you have  an algebraic formalism, 
 a geometry and a one-to-one correspondence that translates the geometry into the algebra and {\em vice versa}. In QM the algebra is perfectly known and 
 validated as exact because it agrees to very high precision with all experimental data. But the meaning of the algebra, i.e. its physical interpretation in terms of a geometry and a dictionary is not  known. This geometrical meaning is provided by the group representation theory itself.
 The point is now, that if you knew that geometry you would discover that some of the algebra is undefined geometrical nonsense. Nonetheless
 this meaningless algebra leads to the correct final result. This
 way it agrees with experimental data, while the geometrical nonsense leads to the paradoxes.
 The geometrical meaning of a spinor is that it is a notation for a group element. 
 Spinors  can {\em a priori} not be  combined linearly as vectors \cite{Coddens} like physicists do, because in the Lorentz and rotation groups
 the operations $g_{2}g_{2}$ are defined, but the operations $c_{1}g_{1} + c_{2} g_{2}$ are not. Spinors belong to a curved manifold, not a vector space.
 As the algebra of QM yields correct results despite this transgression,  a special effort must be made to explain ``why the fluke happens'' by finding 
 {\em a posteriori} a meaning for the algebraic procedure of making linear combinations of spinors. This can be compared  to 
 the way we were forced to
 justify {\em a posteriori}
doing algebra with the quantity $\imath = \sqrt{-1}$ in mathematics, because it brought a wealth of meaningful results.
 It turns then out that we must distinguish between two principles: a true superposition principle (which is physically meaningful)  and a 
 Huyghens' principle (which is physically meaningless but yields excellent numerical results).
 
 The true superposition principle, based on the linearity
 of the equations is that a linear combination $\psi  = \sum_{j} c_{j} \chi_{j}$ is a solution of a Schr\"odinger equation:
 
 \begin{equation} \label{exact}
 -{\hbar^{2}\over{2m}} \Delta \psi + V({\mathbf{r}}) \,\psi= - {\hbar\over{\imath}} {\partial\over{\partial t}} \psi.
  \end{equation}
 
 \noindent when all wave functions $\chi_{j}$ are solutions of the {\em same} Schr\"odinger equation Eq. \ref{exact}. 
 This is then a straightforward mathematical result, and one can argue \cite{Coddens-HAL} that it
 leads to the probability rule $p =  \sum_{j} |c_{j}|^{2} |\chi_{j}|^{2}$ (incoherent summing), whereby one combines probabilities 
 $p_{j} =  |\chi_{j}|^{2}$ in the classical way,
 which corresponds to common sense.
 But telling that the solution $\psi_{1}$ of a first equation with potential $V_{1}$ can be added to the solution $\psi_{2}$ of a second equation
with a different potential $V_{2}$
 to yield a solution $\psi_{3}$ for a third equation with a yet different potential $V_{3}$ can {\em a priori} not be justified by the mathematics 
 and is not exact. It has nothing to do with the linearity of the equations. Summing the equations for $\psi_{1}$ and $\psi_{2}$ does not yield the equation
 for $\psi_{3}$.  A solution of the wave equation for the single-slit
experiment will not necessarily satisfy all the boundary conditions of the double-slit experiment, and vice versa.
 At the best, $\psi_{3}=  \psi_{1} + \psi_{2}$ will in certain physical situations be an excellent approximation. But the fact that this is not rigorously exact
 (in other words {\em logically flawed}, because flawless logic can only yield a result that is rigorously exact)
 and  should be merely considered as a good numerical result rather than an exact physical truth
 is important. In fact,  based on textbook presentations one could  believe that it is an absolute physical truth in principle
 that one must replace the traditional rules of probability calculus $p_{3} = p_{1} + p_{2}$ by substituting the probabilities by their amplitudes.
This is just not true. The belief must be vigorously eradicated because it leads to the misconception that there could exist a deep logical principle
behind $\psi_{3}=  \psi_{1} + \psi_{2}$, that in its proper context would be a truth that is as unshakable as $p_{3} = p_{1} + p_{2}$
in our traditional logic and transcend all human understanding. As discussed below, mistaking the  principle of substituting $p$ by $\psi$ for a deep mysterious absolute truth 
leads to insuperable conceptual problems
in the case of destructive interference where $ \psi_{1}({\mathbf{r}}) + \psi_{2}({\mathbf{r}}) =0$. 
The only way to solve this paradox is following the track of the logical loophole.
What happens here is that the procedure
of adding spinors  is logically  flawed but its result numerically accurate. 
Of course, agreement with experiment  can only validate here  the numerical accuracy, 
not the flawed logic that has been used to obtain it.

To take this objection into account rigorously, we will define that the approximate solution $\psi_{3} \approx \psi_{1} + \psi_{2}$
of the double-slit wave equation follows a Huyghens' principle
and note it as $\psi_{3} =  \psi_{1} \boxplus \psi_{2}$ to remind that it is only numerically accurate, reserving
 the term superposition principle for the case when we combine wave functions that are all solutions of the same linear equation.
 We make this distinction between the superposition principle (with incoherent summing) and a Huyghens' principle (with coherent summing)
 to make sure that we respect what we can do and what we cannot do with spinors. This
 lays also a mathematical basis for justifying that we have two different rules for calculating probabilities and 
 that both the incoherent rule $p =  \sum_{j} |c_{j}|^{2} |\chi_{j}|^{2}$
 and  the coherent rule $p = |\psi|^{2} = | \sum_{j} c_{j} \chi_{j}|^{2}$ are correct within their respective domains of validity.
 This is the mathematical essence of the problem. QM just tells us that once we have an {\em exact} pure-state solution of a wave equation, we must square the
 amplitude of the wave function to obtain an {\em exact} probability distribution.

 The double-slit paradox is so difficult that it has the same destabilizing effect as gaslighting. One starts doubting about one's own mental capabilities.
 But the very last thing we can do in face of such a very hard paradox is to capitulate and think that we are not able to think straight.
 We will thus categorically refuse to yield to such defeatism.
If we believe in logic, the rule $p_{3} = p'_{1} + p'_{2}$, where $p'_{1}$ and $p'_{2}$ are the probabilities to traverse
the slits in the double-slit experiment,  must still be exact. We are then compelled to  conclude that in QM
the probability $p'_{1}$ for traversing slit S$_{1}$ when slit S$_{2}$ is open is  manifestly different
from the probability $p_{1}$ for traversing slit S$_{1}$ when slit S$_{2}$ is closed. We can then ask with Feynman how the particle can ``know''
if the other slit is open or otherwise if its interactions are local.

\section{Local interactions, non-local probabilities} \label{global-local}

The solution to that problem is that {\em the interactions} of the electron with the device {\em are locally defined} 
while the probabilities defined by the wave function are not. {\em The probabilities are non-locally, globally defined.}
When we follow our intuition, the electron interacts with the device in one of the slits. The corresponding probabilities are local interaction probabilities.
We may take this point into consideration. Following our intuition we may  then think that after
doing so we are done. But in QM the story does not end here.  The probabilities are globally defined and
we must solve the wave equation with the global boundary conditions.
We may find locally a solution to the wave equation based on the consideration of the local interactions, but that is not good enough.
The wave equation must also satisfy boundary conditions that are far away from the place where the electron is interacting.
The QM probabilities are defined with respect to the global geometry of the set-up. This global geometry is
fundamentally non-local in the sense that the local interactions of the electron cannot be affected by all aspects of the geometry. Due to this fact 
the ensuing probability distribution is also non-locally defined. 
This claim may look startling. To make sense of it we propose the following slogan, which we will explain below: 
{\em ``We are not studying electrons with the measuring device,
we are studying the measuring device with electrons''.} This slogan introduces a paradigm shift that will 
grow to a leading principle as we go along. We can call it the holographic principle (see below).

In fact, we cannot measure the interference pattern in the 
double-slit experiment with one electron impact on a detector screen. We must make statistics of many electron impacts. We must thus use many electrons and measure
a probability distribution for them.  The probabilities  must be defined in a globally self-consistent way.
The definitions of the probabilities that prevail at one slit may therefore be subject  to compatibility constraints
imposed by the definitions that prevail at the other slit.
We are thus measuring the probability distribution of an ensemble of electrons in interaction with the whole device.
While a single electron cannot ``know'' if the other slit is open or otherwise, the ensemble of electrons
will ``know'' it, because all parts of the measuring device will eventually be explored by the ensemble
of electrons if this ensemble is large enough, {\em i.e.} if our statistics are good enough. When this is the case,
the interference pattern will appear. Reference  \cite{Tonomura} gives actually a nice illustration of how the interference pattern builds up with time.

The geometry of the measuring device is non-local in the sense that a single electron cannot explore  all aspects of the set-up through
its local interactions. There is no contradiction with relativity in the fact that the probabilities for these local interactions
must  fit into a global probability scheme that is dictated also by parts of the set-up a single electron cannot probe. We must thus realize how 
Euclidean geometry contains information that   in essence is non-local, because it cannot all be probed by a single particle, but that
this is not in contradiction with the theory of relativity. The very Lorentz frames used to write down
the Lorentz transformations are non-local because they assume that all clocks in the frame are synchronized up to infinite distance.
It is by no means possible to achieve this,  such that the very tool of a Lorentz frame conceptually violates the theory of relativity. 
But this  remains without any practical incidence on the validity of the theory.

\section{A classical analogy} \label{analogy}

We can render these ideas  clear by an analogy. Imagine a country that sends out spies to an enemy country.
The electrons behave as this  army of spies. The double-slit set-up is the enemy country. 
The physicist is the country that sends out the spies. Each spy is sent to a different part of the
enemy's country, chosen by a random generator. They will all take photographs of the part of the enemy country they end up in. 
The spies may have an action radius of only a kilometer.
Some of the photographs of different spies will overlap. These photographs correspond to the
spots left by the electrons on your detector. If the army of spies you send out is large enough, then
in the end the army  will have made enough photographs 
to assemble  a very detailed complete map of the country. That map corresponds to the
interference pattern. In assembling the global map from the small local patches presented by the photographs we must make sure that the errors do not accumulate such
that everything fits together self-consistently. This is somewhat analogous with the boundary conditions of the wave function that must be satisfied
globally, whereby we can construct the global wave function also by assembling patches of local solutions.
The tool one can use to ensure this global consistency is a Huyghens' principle. An example of such a Huyghens' principle is Feynman's path integral method
or Kirchhoff's method in optics.
The principle is non-local and is therefore responsible for the fact that we must carry
out calculations that are purely mathematical but have no real physical meaning. They may look incomprehensible if we take them literally,
because they may involve e.g. backward propagation in space and even in time  \cite{Feynman4,Cramer}, not to mention photons traveling faster than light.

The interference pattern presents this way the information about the whole experimental set-up.
It does not present this information directly but in an equivalent way, by an integral transform.
This can be seen from Born's treatment  of the scattering of particles of mass $m_{0}$ by a potential $V_{s}$, which  leads to the differential cross section:

\begin{equation} \label{Born}
{d\sigma\over{d\Omega}} = {m_{0}\over{4\pi^{2}}}\, |{\mathscr{F}}(V_{s})({\mathbf{q}}) |^{2},
\end{equation}

\noindent where ${\mathbf{p}} = \hbar{\mathbf{q}}$ is the momentum transfer.
The integral transform is here the Fourier transform ${\mathscr{F}}$, which is a even a one-to-one mapping.
This result is derived within  the Born approximation and is therefore an approximate result. 
In a more rigorous setting, the integral transform could be e.g. the one proposed
by Dirac \cite{Dirac}, which Feynman was able to use to derive the Schr\"odinger equation \cite{Feynman2}. The Huyghens' principles used by Feynman and Kirchhoff
are derived from  integral transforms to which they correspond. (In Feynman's path integral there will be paths that
thread through both slits, which shows that $\psi_{3} = \psi_{1} + \psi_{2}$ is not rigorously exact).
In a double-slit experiment, $V_{s}$ embodies just the geometry of the set-up. 
Combined with a reference beam  ${\mathscr{F}}(V_{s})({\mathbf{q}})$ yields its hologram.
The spies in our analogy are not correlated and not interacting, but the information about the country
is correlated: It is the information we put on a map. The map will e.g. show correlations in the form of long straight lines, roads that stretch out for thousands of miles,
but none of your spies  will have seen these correlations
and the global picture. They just have seen the local picture of the things that were situated within their action radius. 
The global picture, the global information about the enemy country is non-local, and contains correlations,
but it  can nevertheless be obtained if you send out enough spies to explore the whole country, and it will show on the map assembled. 
That is what we are aiming at by  invoking  the non-locality of the Lorentz frame and the non-locality
of the wave function.
The global information gathered by
many electrons contains the information how many slits are open. It is that kind of global information
 about your set-up that is contained in the wave function.
You need many single electrons to collect that global information. 
A single electron just gives you one
impact on the detector screen. That is almost no information. Such an impact is a Dirac delta measure, derived from the Fourier transform
of a flat distribution. It contains hardly any information about the set-up because it does not provide any contrast.
This global geometry contains thus more information than any single electron can measure through its local interactions.
And it is here that the paradox creeps in. The probabilities are not defined locally, but globally.
The interactions are local and in following our intuition,  we infer from this that the definitions of the probabilities will be local as well,
but they are not. The space wherein the electrons  travel in the double-slit experiment is not simply connected, 
which is, as we will see, a piece of global, topological information  apt
to profoundly upset the way we must define probabilities.

\section{Highly simplified descriptions still catch the essence} \label{no-theory}

The description of the experimental set-up  we use to calculate a wave function is conventionally highly idealized and simplified.
Writing an equation that would make it possible to take into account all atoms of the macroscopic device in the experimental
set-up is a hopeless task. Moreover, the total number of atoms in ``identical'' experimental set-ups is only approximately identical.
In such a description there is  no thought for the question  if the local interaction of a neutron  involves a spin flip or otherwise. 
Despite its crudeness, such a purely geometrical description is apt to seize
a crucial ingredient of any experiment whereby interference occurs. It is able to account for the difference between set-ups with one and two slits, 
as in solving the wave equation we unwittingly avoid the pitfall of ignoring the difference
between globally and locally defined probabilities, rendering the solution  adopted  tacitly global.
 In this sense the probability paradox we are confronted with is akin to Bertrand's paradox in probability calculus.
 It is not sufficient to calculate the interaction probabilities locally. We must further specify how we will use these probabilities later on in the procedure to fit
them  into a global picture. The probabilities will be only unambiguously defined if we define simultaneously the whole protocol  we will use to calculate with them.

\section{Winnowing out the over-interpretations} \label{exit-Bohr}

It is now time to get rid of the particle-wave duality.
 Electrons are always particles, never waves. 
As pointed out by Feynman, electrons are always particles
because a detector detects always a full electron at a time, never a fraction of an electron. 
Electrons never travel like a wave through both slits simultaneously. But in a sense, their probability distribution does.
It is the probability distribution of many electrons which displays wave behaviour and acts like a flowing liquid, 
not the individual electrons themselves.
This postulate only reflects literally what QM says, viz. that the wave function is a probability amplitude,
and that it behaves like a wave because it is obtained as the solution of a wave equation. 
Measuring the probabilities requires measuring many electrons, such that the probability amplitude is a probability amplitude defined by considering
an ensemble of electrons  \cite{Ballentine} with an ensemble of possible histories.
Although this sharp dichotomy is very clearly present
in the rules, we seem to loose sight of it when we are reasoning intuitively. This is due to a tendency
towards {\em ``Hineininterpretierung''}  in terms of  Broglie's initial idea that the particles themselves, not their probability distributions,
would be waves.
These heuristics have historically been useful but are reading more into the issue than there really is. Their addition
blurs again the very accurate sharp pictures provided by  QM. With hindsight, we must therefore 
dispense with the particle-wave duality. The rules of QM are clear enough in their own right: {\em In claro non interpretatur!}
Wave functions also very obviously do not collapse. They serve to describe a statistical ensemble of possible events, not outcomes of  single events.

It is also time to kill the traditional reading of $\psi_{3} = \psi_{1} \boxplus \psi_{2}$ in terms of a ``superposition principle'', based on the wave picture.
 It is is  only  a convenient numerical  recipe, a Huyghens' principle without true physical meaning. We can make the experiment in such a way
that only one electron is emitted by the source every quarter of an hour. Still the interference pattern will build up if we wait long enough. 
But if $\psi_{1}$ and $\psi_{2}$ were to describe the correct probabilities from slit S$_{1}$ and slit S$_{2}$, we would never be able to explain destructive interference.
How could a second electron that travels through slit S$_{2}$ erase the impact made on the detector screen of an electron that traveled through
slit S$_{1}$ hours earlier? We may speculate that the electron feels whether  the other slit is open or otherwise. E.g. the electron might polarize
the charge distribution inside the measuring device and the presence of the other slit might influence this induced charge distribution.
This would be an influence at a distance that is not incompatible with the theory of relativity. But this scenario is not very likely. As pointed out by Feynman 
 interference  is  a universal phenomenon. It exists also for photons, neutrons, helium atoms, etc... 
 We already capture the essence of this universal phenomenon in a simple, crude geometrical description of the macroscopic set-up of the experiment.
While this could be a matter of pure luck according to the principle that fortune favors fools,
it is not likely that one could translate the scenario evoked for electrons  to an equivalent scenario in all these
 different situations.
E.g. how could the fact that another slit is open (in a nm-sized double-slit experiment) affect the nuclear process at the fm scale of the spin flip of a neutron? 
The generality of the scenario based on an influence at a distance is thus not very likely.

We must thus conclude that  $\psi_{3} = \psi_{1} \boxplus \psi_{2}$ is a very good numerical approximation for the true wave function  $\psi'_{3}$,
whereby the physically meaningful identity reads $\psi'_{3} = \psi'_{1} + \psi'_{2}$
 in terms of other wave functions $ \psi'_{1}$ and $\psi'_{2}$. 
 The wave functions $ \psi'_{1}$ and $\psi'_{2}$ must now both be zero, $\psi_{1}' ({\mathbf{r}})= \psi_{2}' ({\mathbf{r}})= 0$, 
  in all places ${\mathbf{r}}$ where we have ``destructive interference'', because $p_{3} = p_{1} + p_{2}$  must still be valid.
  In other words $\psi_{1} \neq \psi'_{1}$ and $\psi_{2} \neq \psi'_{2}$.

\section{Undecidability} \label{ternary-logic}
 
We can further improve our intuition for this by another approach that 
addresses more the way we study electrons with the set-up and is
based on undecidablity.
The concept of undecidablity has been formalized in mathematics, which provides many examples of
 undecidable questions.
Examples occur e.g. in G\"odel's theorem  \cite{Godel}.
The existence of such undecidable questions may look hilarious to common sense but this does not need to be.
In fact, the reason for the existence of such undecidable questions
is that the set of axioms of the theory is incomplete. We can complete then the theory by adding an axiom telling the answer to the question
is ``yes'', or by adding an axiom telling the answer to the question is ``no''. The two alternatives permit to stay within a system based on binary logic
({\em ``tertium non datur''}) and  lead to two different axiomatic systems and
thus to two different theories. An example of this are Euclidean and hyperbolic geometry \cite{hyperbolic}. In Euclidean geometry one has added
on the fifth parallels postulate to the first four postulates of Euclid, 
while in hyperbolic geometry one has added on an alternative postulate that is at variance with the parallels postulate. 
We are actually not forced to make a choice: We can decide to study a ``pre-geometry'', wherein the question remains undecidable.
The axiom one has to add can
be considered as information that was lacking in the initial set of four axioms. Without adding it one cannot address  the yes-or-no question which reveals
that  the axiomatic system without the parallels postulate added  is incomplete. As G\"odel has shown, we will almost always run eventually  into such a problem of
incompleteness. On the basis of Poincar\'e's mapping between hyperbolic and Euclidean geometry \cite{hyperbolic}, we can appreciate which information  was lacking
in the first four postulates. The information was not enough to identify the straight lines as really straight, as  we could still interpret the straight lines in terms of half circles in a half plane.

When the interactions are coherent in the double-slit experiment, the question through which one of the two slits the electron has traveled
is very obviously also experimentally undecidable. Just like in mathematics, this is due to lack of information. We just do not have the information
that could permit us telling which way the electron has gone. This is exactly what Feynman pointed out so carefully.
In his lecture he considers three possibilities for our observation of the history of an electron: ``slit S$_{1}$'', ``slit S$_{2}$'', and ``not seen''. 
 The third option corresponds exactly to this concept of undecidability. 
He works this  out with many examples in reference \cite{Feynman1}, 
to show that there is a one-to-one correspondence between undecidability and coherence.
Coherence already occurs in a single-slit experiment, where it is at the origin of the diffraction fringes. But in the double-slit experiment
the lack of knowledge becomes all at once amplified to an objective undecidability of the question through which slit the electron has traveled, which does not exist in the single-slit experiment.
What happens here in the required change of the definition of the probabilities has nothing to do with a change in local physical interactions.
It has only to do with the question how we define a probability with respect to a body of available information.
The probabilities are in a sense conditional because they depend on the information available.
As the lack of information is different in the double-slit experiment, the body of information available changes, 
such that the probabilities must be defined in a completely different way (Bertrand's paradox).
Information biases probabilities, which is why insurance companies ask their clients to fill forms requesting information about them.

We have methods to deal with such bias.
According  to common-sense intuition whereby we reason only on the local interactions, opening or closing the other slit
 would not affect the probabilities or only affect them slightly, but this is wrong. 
 We may also think that the undecidability is just experimental
 such that it would not matter for performing
our probability calculus. We may reckon that in reality, the electron 
must have gone through one of the two slits anyway.
We argue then that we can just assume that half of the electrons went one way, and the other half of the electrons the other way, and that
we can then use statistical averaging to simulate the reality, just like we do in classical statistical physics to remove bias. 
We can verify this argument by detailed QM calculations. We can calculate
 the solutions of the three wave equations
in Eq. \ref{slit1} and compare $|\psi_{3}|^{2}$ with the result of our averaging procedure based on  $|\psi_{1}|^{2}$ and  $|\psi_{2}|^{2}$.
This will reproduce  the disagreement between the experimental data and our classical intuition, confirming QM is right.

To make sense of this we may argue that we are not used to logic that allows for undecidability. 
Decided histories with labels S$_{1}$ or S$_{2}$ occur in a theory based on a system of axioms ${\cal{A}}_{1}$ (binary logic),
while the undecided histories occur in a theory based on an all together different system of axioms ${\cal{A}}_{2}$ (ternary logic).
In fact, the averaging procedure is still correct in ${\cal{A}}_{2}$ because the electron travels indeed either through S$_{1}$ or  through S$_{2}$ following
binary logic. But the information we obtain about the electron's path does not follow  binary logic. It follows ternary logic.

Due to the information bias the probabilities $|\psi'_{1}|^{2}$ and  $|\psi'_{2}|^{2}$ to be used  in 
${\cal{A}}_{2}$ are very different from the probabilities $|\psi_{1}|^{2}$ and  $|\psi_{2}|^{2}$ to be used in ${\cal{A}}_{1}$.
The paradox results thus from the fact that we just did not imagine that such a difference could exist.
Assuming $\psi'_{j} =  \psi_{j}$, for $j=1,2$ amounts to neglecting the ternary bias of the information contained in our data 
and reflects the fact that we are not aware of the global character of the definition of the probabilities. 
To show  that the intuition  $\psi'_{j} =  \psi_{j}$, for $j=1,2$ is wrong, nothing is better than giving a counterexample. The counterexample is the double-slit experiment where clearly
the probability is not given by $p_{3} = |\psi_{1}|^{2} + |\psi_{2}|^{2}$ but by $p_{3} = |\psi'_{1}|^{2} + |\psi'_{2}|^{2} \approx |\psi_{1} \boxplus \psi_{2}|^{2}$, where the 
index 3 really refers to the third (undecidable) option. 
It is then useless to insist any further.

 The undecidability criterion corresponds to a global constraint  that has a spectacular  impact on the definition of the probabilities. The probabilities are {\em conditional}
and {\em not absolute}. They are {\em physically} defined by
the physical information gathered from the interactions with the set-up,  {\em not absolutely} 
by some absolute divine knowledge  
about the path the electron has taken. 
The set-up biases the information we can obtain about that divine knowledge by withholding a part of the information about it.
Einstein is perfectly right that the Moon is still out there when we are not watching.
But we cannot find out that the Moon is there if we do not register any of its interactions with its environment,  even if it is there.  
If we do not register any information about  the existence of the Moon,
then the information contained in our experimental results must be biased in such a way that everything looks as though the Moon were not there \cite{Spekkens}. 
Therefore, in QM the undecidability must affect the definition of the probabilities and bias them, such that $p'_{j} \neq p_{j}$, for $j=1,2$. 
The experimental probabilities  must reflect the undecidability.
In a rigorous formulation,  this undecidability becomes a consequence of the fact that the wave function must be a function, because it is the integral transform
of the potential, which must represent all the information about the set-up and its built-in undecidability. As the phase of the wave function corresponds to the 
 spin angle of the electron, even this angle is thus uniquely defined. 
 
 \section{The correct analysis of the experiment} \label{making-it-right}

This idea is worked out in reference  \cite{Coddens}, pp. 329-333,
and depends critically on the fact that the space traversed by the electrons that end up in the detector is not simply connected.
It is based on the simplifying ansatz that the way the electron travels through the set-up from a point ${\mathbf{r}}_{1}$
 to a point ${\mathbf{r}}_{2}$ has
 no incidence whatsoever on the phase difference  of the wave function between ${\mathbf{r}}_{1}$
and ${\mathbf{r}}_{2}$. 
The idea is based on an Aharonov-Bohm type of argument: For two alternative paths $\Gamma_{1}$ and $\Gamma_{2}$  between ${\mathbf{r}}_{1}$
and ${\mathbf{r}}_{2}$, we have
$[\, \int_{\Gamma_{1}} Edt - {\mathbf{p\cdot}}d{\mathbf{r}}\,] - [\, \int_{\Gamma_{2}} Edt - {\mathbf{p\cdot}}d{\mathbf{r}}\,] = 2\pi n $, where
 $n\in {\mathbb{Z}}$. The union of the two paths defines a loop. In a single-slit experiment this loop can be shrunk continuously to
point which can be used to prove that $n=0$. In a double-slit experiment the loop cannot be shrunk to a point when
$\Gamma_{1}$ and $\Gamma_{2}$ are threading through different slits, such that  $n \neq 0$ becomes then possible. 
Each interference fringe corresponds to one value of $n\in {\mathbb{Z}}$. 
A phase difference of  $2\pi n $ occurs also in the textbook approach where one argues that to obtain constructive interference 
the difference in path lengths behind the slits
must yield  a phase difference $2\pi n$. But this resemblance does not run deep and is superficial. 
The textbook approach deals with phase differences between $\psi_{1}$
and $\psi_{2}$ in special points ${\mathbf{r}}_{2}$, while our approach deals with different phases built up over paths $\Gamma_{1}$ and $\Gamma_{2}$
within $\psi_{3} = \psi_{1} \boxplus \psi_{2}$ for all points ${\mathbf{r}}_{2}$.

 We can approach this somewhat differently.  We will show that the textbook quantum mechanics prescription $\psi_{3} = \psi_{1} \boxplus \psi_{2}$
belongs to ``pre-geometry'' in the analogy we discussed above. We accept that the question through which slit the electron has traveled is undecidable
and accept ternary logic.
We should then play the game and not attempt in any instance to reason about the question which way the electron has traveled,
because this information is not available. But we can also add a new axiom, the axiom of the existence of a divine perspective, rendering the question decidable
for a divine observer who can  also observe the information withheld by the set-up.
We must then also play the game and accept the fact that the probabilities we will discuss can no longer be measured, such
that the conclusions we draw will now no longer be compelled by experimental evidence but by the pure binary logic imposed by the addition of the axiom.
 We take the exact solution $\psi'_{3}$ of the double-slit experiment
 and try to determine the parts $\psi'_{1}$ and $\psi'_{2}$ of it that stem from slits S$_{1}$ and S$_{2}$. We can mentally imagine such
 a partition without making a logical error because each electron must go through one of the slits, even if we will never know which one.
  We have thus $\psi'_{3} = \psi'_{1} + \psi'_{2}$. 
We expect that $\psi'_{1}({\mathbf{r}})$  must vanish  on slit S$_{2}$ and  $\psi'_{2}({\mathbf{r}})$  on slit S$_{1}$, but  
  we must refrain here from jumping to conclusions by deciding
   that $\psi'_{1} = \psi_{1}$ and $\psi'_{2} = \psi_{2}$.
 We can only attribute probabilities $|\psi'_{1}|^{2}$ and  $|\psi'_{2}|^{2}$ to the slits, based on the lack of experimental knowledge.
We will use
$\psi'_{3} = \psi_{3} = \psi_{1} \boxplus \psi_{2}$ in our calculations, as we know it is a good numerical approximation.
Let us call the part of ${\mathbb{R}}^{3}$ behind the slits $V$. 
Following the idea that $\psi'_{1}({\mathbf{r}})$  would have to vanish  on slit S$_{2}$ and  $\psi'_{2}({\mathbf{r}})$  on slit S$_{1}$,
we subdivide $V$ in a region $Z_{1}$ where $\psi'_{1}({\mathbf{r}}) =0$ and a region $N_{1}$ where $\psi'_{1}({\mathbf{r}}) \neq 0$.
We define $Z_{2}$ and $N_{2}$ similarly. In the region $N_{1}\cap Z_{2}$ we can multiply $\psi'_{1}$ by an arbitrary phase $e^{\imath \chi_{1}}$
without changing $|\psi'_{1}({\mathbf{r}})|^{2}$. In the region $Z_{1}$ this is true as well as $|\psi'_{1}({\mathbf{r}})|^{2} =0$.
Similarly $\psi'_{2}({\mathbf{r}})$ can be multiplied by an an arbitrary phase $e^{\imath \chi_{2}}$ in the regions
$ Z_{2}$ and $Z_{1}\cap N_{2}$. Let us now address the region $N_{1}\cap N_{2}$.
We must  certainly have $|\psi'_{1}({\mathbf{r}})|^{2} +  |\psi'_{2}({\mathbf{r}})|^{2} = |\psi_{3}({\mathbf{r}})|^{2}$, because the probabilities
 for going through  slit S$_{1}$ and  for going through  slit S$_{2}$ are mutually exclusive and must add up to the total probability of transmission.
 We might have started to doubt about the correctness of this idea, due to the way textbooks present the problem, but we should never have doubted.
 Let us   thus put
  $\psi'_{1}({\mathbf{r}}) = |\psi_{3}({\mathbf{r}})| \cos\alpha \,e^{\imath \alpha_{1}}$,
$\psi'_{2}({\mathbf{r}}) = |\psi_{3}({\mathbf{r}})| \sin\alpha \,e^{\imath \alpha_{2}}$, $\forall {\mathbf{r}} \in N_{1} \cap N_{2} =W$. 
In fact, if $\psi'_{j}$ is a partial solution for the slit S$_j$,
$\psi'_{j} e^{\imath \alpha_{j}}$ will also be a partial solution for the slit S$_j$. 
We must take here $\alpha$, $\alpha_{1}$ and $\alpha_{2}$ as constants.
If we took a solution whereby $\alpha$, $\alpha_{1}$ and $\alpha_{2}$ were functions
of ${\mathbf{r}}$,  the result obtained would no longer be a solution of the Schr\"odinger equation in free space, 
due to the terms containing the spatial derivatives of  $\alpha$, $\alpha_{1} $ and $\alpha_{2} $ which are not zero.
In first instance this argument shows also that we must take $\chi_{1} = \alpha_{1}$ and $\chi_{2} = \alpha_{2}$. 
We do not have to care about the
phases $\chi_{1}$ and $\chi_{2}$ in the single-slit experiments, but this changes  in the double-slit experiment, because
we must make things work out globally. Over $N_{1}\cap Z_{2} $, we must have  
$\psi'_{1}({\mathbf{r}})= \psi_{3}({\mathbf{r}}) = \psi_{1}({\mathbf{r}}) $, as $\psi_{2}({\mathbf{r}})  = 0$.
Similarly, $(\,\forall {\mathbf{r}} \in N_{2}\cap Z_{1}\,)\,  (\,\psi'_{2}({\mathbf{r}}) = \psi_{3}({\mathbf{r}}) = \psi_{2}({\mathbf{r}})\,) $  as $\psi_{1}({\mathbf{r}})  = 0$.
Then $\int_{W}\, |\psi'_{1}({\mathbf{r}})|^{2}\,d{\mathbf{r}} = \cos^{2}\alpha \int_{W}\, |\psi'_{3}({\mathbf{r}})|^{2} \,d{\mathbf{r}}$ 
and $\int_{W}\, |\psi'_{2}({\mathbf{r}})|^{2}\,d{\mathbf{r}} = \sin^{2}\alpha \int_{W}\, |\psi'_{3}({\mathbf{r}})|^{2}\,d{\mathbf{r}} $.
Due to the symmetry, we must have $\int_{W}\, |\psi'_{2}({\mathbf{r}})|^{2}\,d{\mathbf{r}}  = \int_{W}\, |\psi'_{1}({\mathbf{r}})|^{2}\,d{\mathbf{r}} $,
such that $\alpha = {\pi\over{4}}$. We see from this that not only $\int_{W}\, |\psi'_{2}({\mathbf{r}})|^{2}\,d{\mathbf{r}}  = \int_{W}\, |\psi'_{1}({\mathbf{r}})|^{2}\,d{\mathbf{r}} = {1\over{2}}
 \int_{W}\, |\psi'_{3}({\mathbf{r}})|^{2} \,d{\mathbf{r}}$, but also $ |\psi'_{2}({\mathbf{r}})|^{2} =  |\psi'_{1}({\mathbf{r}})|^{2} = {1\over{2}}  |\psi'_{3}({\mathbf{r}})|^{2}$.
 In each point ${\mathbf{r}} \in W$ the probability
that the electron has traveled through a slit to get to ${\mathbf{r}}$  is equal to the probability that it has traveled through the other slit.
This is due to the undecidability.
The choices  $\alpha_{1} \neq 0$, $\alpha_{2}\neq 0$ we have to impose on the phases,
 are embodying here  the idea that a solution
of a Schr\"odinger equation with potential $V_{j}$, for $j=1,2$ cannot be considered as a solution of a Schr\"odinger equation with potential $V_{3}$.
The conditions we have to impose on  $\alpha_{1}$ and $\alpha_{2}$ are thus a kind of disguised boundary conditions. They are not true boundary boundaries,
but a supplementary condition (a logical constraint) that we want $\psi'_{1}$ and $\psi'_{2}$ to obey  ``divine''  binary logic.

We can summarize these results
as $\psi'_{1} = {1\over{\sqrt{2}}} |\psi_{1} \boxplus \psi_{2}| \,e^{\imath \alpha_{1}}$ and
$\psi'_{2} = {1\over{\sqrt{2}}} |\psi_{1} \boxplus \psi_{2}| \, e^{\imath \alpha_{2}}$. 
Let us write
$\psi_{1} \boxplus \psi_{2} = |\psi_{1} \boxplus \psi_{2}| \, e^{\imath \chi}$.
We can now calculate
$\alpha_{1}$ and $\alpha_{2}$ by identification. This yields on $W$:

\begin{equation} \label{split}
\begin{tabular}{ccccc}
$\psi'_{1}$ & $=$ & $ {|\psi_{1} \boxplus \psi_{2}| \,e^{\imath(\chi + {\pi\over{4}})}\over{\sqrt{2}}}$ & 
$=$ & ${1\over{\sqrt{2}}}\,(\psi_{1} \boxplus \psi_{2})\,e^{+ \imath  {\pi\over{4}}}$ 
$\neq$  $ \psi_{1}$ \\
&&&&\\
$\psi'_{2}$ & $=$ & ${|\psi_{1} \boxplus \psi_{2}| \, e^{\imath(\chi - {\pi\over{4}})}\over{\sqrt{2}}}$ & 
$=$ & ${1\over{\sqrt{2}}}\,(\psi_{1} \boxplus \psi_{2})\,e^{- \imath  {\pi\over{4}}}$ 
$ \neq$ $ \psi_{2}$ \\
\end{tabular}
\end{equation}

\noindent What this shows is that the rule $\psi_{3} = \psi_{1} \boxplus \psi_{2}$ is  logically 
flawed, because the correct expression is  $ \psi'_{3} = \psi'_{1} + \psi'_{2}$. We were able
to get an inkling of this loophole by noticing that $\psi_{3} = \psi_{1} \boxplus \psi_{2}$ is not rigorously exact, even if it is an excellent approximation.
 The differences between $\psi'_{j}$ and $\psi_{j}$, for $j=1,2$, are not negligible.
The phases of  $\psi'_{1}$ and $\psi'_{2}$ always differ by ${\pi\over{2}}$ such that they are fully correlated.
The difference between the phases of $\psi_{1}$ and $\psi_{2}$ can be anything. They can be opposite (destructive interference) or identical
(constructive interference).
In fact, in contrast to $\psi_{1}$ and $\psi_{2}$, $\psi'_{1}$ and $\psi'_{2}$ reproduce the oscillations of the interference pattern.
In this respect, the fact that $\alpha_{1}$ and $ \alpha_{2}$ are different by a fixed amount is crucial.
It permits to make up for the normalization factor ${1\over{\sqrt{2}}}$ and end up with the correct numerical result
of the flawed calculation $\psi_{1} \boxplus \psi_{2}$.
The  phases of $\psi'_{1}$ and $\psi'_{2}$  conspire to render
$\psi'_{1} + \psi'_{2}$ equal to $\psi_{1} \boxplus \psi_{2}$.

However, at the boundaries  of $N_{1}\cap Z_{2}$ and $N_{2}\cap Z_{1}$ with $W$ there are awkward discontinuities.
In $N_{1}\cap Z_{2}$, we must have $\psi'_{1}({\mathbf{r}}) = \psi_{1}({\mathbf{r}})$, while in $N_{1}\cap N_{2}$, we have
$\psi'_{1}({\mathbf{r}}) = {1\over{\sqrt{2}}}\,(\psi_{1}({\mathbf{r}}) \boxplus \psi_{2}({\mathbf{r}}))\,e^{+ \imath  {\pi\over{4}}}$.
The difference between the solutions is ${\psi_{3}({\mathbf{r}}) \over{\sqrt{2}}}\,e^{\imath(\chi-{\pi\over{4}})}$. This is
the value $\psi'_{2}({\mathbf{r}})$ would take over $N_{1}\cap Z_{2}$ if we extrapolated it from $W$ to $N_{1}\cap Z_{2}$.
We can consider that we can accept this discontinuity at the boundary, because over $N_{1}\cap Z_{2}$, the question through which  slit
the electron has traveled is decidable, while over $W$ it is undecidable, such that there is an abrupt change of logical regime at this boundary.
In reality, the boundary between $W$ and $N_{1}\cap Z_{2}$ could be more diffuse and be the result of an integration over the
slits, such that the description is schematic and the abruptness  not real. The main aim of our calculation is to obtain a qualitative understanding rather than a completely rigorous solution.
The same arguments can be repeated at the boundary between $N_{2}\cap Z_{1}$ and $W$. If we accept this solution, then
$\psi'_{1}({\mathbf{r}})$  will vanish  on slit S$_{2}$ and  $\psi'_{2}({\mathbf{r}})$  will vanish on slit S$_{1}$. \cite{note0}
We  can also consider these discontinuities as a serious issue.
We could then postulate that we must  assume that $N_{1}\cap Z_{2} = \emptyset$ $\&$
$N_{1}\cap Z_{2} = \emptyset$, in order to avoid the discontinuities. 
The fact that we have to choose $N_{1}\cap Z_{2} = \emptyset$ $\&$
$N_{1}\cap Z_{2} = \emptyset$ would then be a poignant illustration of the possible consequences of undecidability.  Contrary to intuition, the value
we  have to attribute  in a point of slit S$_{1}$, to the probability that the particle has traveled through
slit S$_{2}$ is now not zero as we might  have expected  but $ {1\over{2}}  |\psi'_{3}({\mathbf{r}})|^{2}$.
The experimental undecidability biases thus the probabilities such that they are no longer
 the  ``divine probabilities''. The two different approaches  correspond to Einstein-like and Bohr-like viewpoints.
Both approaches are logically tenable when the detector screen is
completely in the zone $W$, because the quantities in  $N_{1}\cap Z_{2}$ and  $N_{2}\cap Z_{1}$ are then not measured quantities.
Of course we could try to measure them by putting the detector screen close to the slits, but this would be a different experiment, leading to different
probabilities, as Feynman has pointed out
in his analysis.

In analyzing the path integrals, one should recover in principle the same results.
However, the pitfall is here that one might
 too quickly conclude that $\psi'_{j} = \psi_{j}$ like in the textbook presentations, which leads us straight into the paradox.
We see thus that the Huyghens' principle is a purely numerical recipe that is physically meaningless, because
it searches for a correct global solution without caring about the correctness of the partial solutions. It follows the experimental ternary logic
and therefore is allowed to mistreat the phase difference that exists between the
partial solutions $\psi'_{1}$ and $\psi'_{2}$ which  always have the same phase difference, such that they cannot interfere
destructively. 
The rule $\psi_{3} = \psi_{1} \boxplus \psi_{2}$ is perfectly right in ternary logic where we decide that we do not bother which way
the particle has travelled, because that question is empirically undecidable. It is then empirically meaningless to separate $\psi'_{3}$ into two parts.
This corresponds to Bohr's viewpoint. It is tenable because it will not be contradicted by experiment.
This changes if one wants to impose also binary logic on the wave function, arguing that conceptually the question about the slits should be decidable
from the perspective of a divine observer. We do need then a correct
decomposition  $\psi'_{3} =  \psi'_{1}  + \psi'_{2}$. We find then out that $\psi'_{j} \neq \psi_{j}$ and we can attribute this
change between the single-slit and the double-slit probabilities to the difference between the ways we must define probabilities in both types of logic.
If we were able by divine knowledge to assign to each electron impact on the detector the corresponding number of  the slit through which the electron has traveled,
we would recover the experimental frequencies $|\psi'_{j}|^{2}$. This is Einstein's viewpoint.
 Having made this clear, everybody is free to decide  for himself if he prefers to study geometry in binary logic or pre-geometry
in ternary logic.
But refusing Einstein's binary logic based on the argument that $\psi'_{1}$ and  $\psi'_{2}$ cannot be measured 
appears to us a stronger and more frustrating
{\em Ansatz} than the one that consists in  introducing variables that cannot be measured. The refusal  is of course in direct line
with Heisenberg's initial program of removing all quantities that cannot be measured from the theory. It is Heisenberg's minimalism which
preserves the experimental undecidability and ternary logic within the theory. To this we can add a supplementary logical constraint which enforces
binary logic.
As Eq. \ref{split} shows, the difference between the  fake  partial ternary solutions $\psi_{j}$ and the correct  partial ternary solutions $\psi'_{j}$ (where $j=1,2$) 
is much larger than we ever might have expected on the basis of the logical loophole that  $\psi_{3} = \psi_{1} \boxplus \psi_{2}$ is not rigorously exact.
In the approach with the additional binary constraint, whereby we follow  the spirit of Bohr, we even do not reproduce $\psi'_{1}({\mathbf{r}}) = 0$ on slit S$_{2}$ and
$\psi'_{2}({\mathbf{r}}) = 0$ on slit S$_{1}$, because these quantities are not measured if we assume that they are only measured far behind the slits.
In the Bohr-like approach, the conditions $\psi'_{1}({\mathbf{r}}) = 0$ on slit S$_{2}$ and
$\psi'_{2}({\mathbf{r}}) = 0$ on slit S$_{1}$ are thus not correct boundary conditions for a measurement far behind the slits, because
they violate the {\em Ansatz} of experimental undecidability. The bias in the experimentally measured probabilities due to the undecidability
cannot be removed by the divine knowledge about the history.
If we want a set-up with the boundary conditions that correspond to the unbiased case whereby $\psi'_{1}({\mathbf{r}}) = 0$ on slit S$_{2}$ and
$\psi'_{2}({\mathbf{r}}) = 0$ on slit S$_{1}$, we must assume that the detector screen is put immediately behind the slits, and the interference pattern can then not be measured, while everything becomes experimentally decidable. 
Otherwise, we must assume that  $\psi'_{1}({\mathbf{r}})$ and $\psi'_{2}({\mathbf{r}})$ are not measured on the slits such that they can
satisfy the undecidable solution in Eq. \ref{split}. The partial probabilities given by $\psi'_{1}({\mathbf{r}})$ and $\psi'_{2}({\mathbf{r}})$
are thus extrapolated quantities, and they are only valid within one set-up with a well-defined position of the detector screen.
Even in the pure Heisenberg approach whereby one postulates that we are not allowed to ask through which slit the electron has traveled, the wave function contains extrapolated quantities that are not measured, despite the original agenda of that approach.
We see also that there is {\em always} an additional  logical constraint that must be added in order to account for
the fact that the options of traveling  through the slits S$_{1}$ or S$_{2}$ are mutually exclusive. This has been systematically overlooked, with
 the consequence that one obtains
the result $p\neq p_{1} + p_{2}$, which
is impossible to make sense of.  It is certainly not justified to use $p\neq p_{1} + p_{2}$ as a starting basis for raising philosophical issues.
Moreover, imposing the boundary condition that  $\psi'_{1}({\mathbf{r}}) = 0$ on slit S$_{2}$ and
$\psi'_{2}({\mathbf{r}}) = 0$ on slit S$_{1}$ remains a matter of  choice, depending on which vision one wants to follow. 
If we do not clearly point out  this choice, then confusion can enter the scene and lead
to paradoxes,  because adding this boundary condition amounts to adding information
and information biases the definition of the probabilities.

In summary,  $\psi_{3} = \psi_{1} \boxplus \psi_{2}$ is wrong if we cheat by wanting to satisfy also  binary logic in the analysis of an experiment
that follows ternary logic by attributing meaning to $\psi_{1}$ and $\psi_{2}$. But
it yields the correct numerical result for the total wave function if we play the game and
 respect the empirical undecidability by not asking which way the particle has traveled. We can thus only
uphold that the textbook rule  $\psi_{3} = \psi_{1} \boxplus \psi_{2}$ is correct if we accept that the double-slit experiment experimentally follows  ternary logic.
Within binary logic, the agreement of the numerical result
with the  experimental data is misleading, as such an agreement does not provide a watertight proof for the correctness of a theory. 
If a theory contains logical and mathematical flaws, then
it must be wrong despite its agreement with experimental data \cite{note}. 

Eq. \ref{split} shows that interference does not exist, because the phase factors $e^{\imath {\pi\over{4}}}$  and $e^{-\imath {\pi\over{4}}}$
of $\psi'_{1}$ and $\psi'_{2}$
always add up
to $\sqrt{2}$. We may note in this respect that $\chi$ itself is only determined up to an arbitrary
constant within the experiment. The wave function can thus not become zero due to phase differences between $\psi'_{1}$ and
 $\psi'_{2}$ like happens with
$\psi_{1}$ and $\psi_{2}$ in $ \psi_{1} \boxplus \psi_{2}$. When $\psi'_{3}$ is zero, both $\psi'_{1}$ and $\psi'_{2}$ are zero.
Interference thus only exists within the purely numerical, virtual reality of the Huyghen's
principle, which is not a  narrative of the real world. We must thus not only dispose of the particle-wave duality, but also  be very wary of the wave pictures
we build based on the intuition we gain from experiments in water tanks.
These pictures are apt to conjure up a very misleading imagery that leads to fake conceptual problems and stirs a lot of confusion.
The phase of the wave function has physical meaning, and spinors can only be added meaningfully
if we get their phases right.

While this solves the probability paradox, we may still ask also for a better understanding of the reasons why  the interference pattern occurs in ternary logic.
In fact, up to now, we have only discussed the phases. We must also discuss the amplitudes of the wave functions.
Let us observe in this respect that  $\psi_{1}^{'*}({\mathbf{r}})\psi'_{2}({\mathbf{r}}) + \psi'_{1}({\mathbf{r}})\psi_{2}^{'*}({\mathbf{r}}) =0,
\forall {\mathbf{r}} \in V$, such that $\psi'_{1}({\mathbf{r}})$ and $\psi'_{2}({\mathbf{r}})$ are everywhere in $V$ orthogonal
with respect to the Hermitian norm. This result actually ensures that $|\psi'_{1} + \psi'_{2} |^{2} =
|\psi'_{1} |^{2} + | \psi'_{2} |^{2}$ such $\psi'_{1}$ and $\psi'_{2}$ are describing mutually exclusive probabilities. To obtain this orthogonality condition
we must have actually that $\psi'_{1} = \psi'_{2}e^{\pm \imath {\pi\over{2}}}$. When this condition is fulfilled,
and $\psi'_{1}$ is zero in slit S$_{2}$ and $\psi'_{2}$ is zero in slit S$_{1}$
the functions $\psi'_{1}$ and $\psi'_{2}$ become exact wave functions for the double-slit experiment.  
They can then actually be summed according the superposition principle. Let us now assume that there exists a wave function 
$\zeta$ for the double-slit experiment. We do not assume here that we know that  $\zeta = \psi_{1} \boxplus \psi_{2}$ must be true, such that we do not know
that it corresponds to an interference pattern.
But it must be possible to decompose it into unknown functions $\zeta_{1}$ and $\zeta_{2}$ according to binary logic as described
for $\psi'_{1}$ and $\psi'_{2}$ above.
We can consider a continuous sweep over  the detector screen from the left to the right, whereby we are visiting the points P. 
Let us call the source S, and the centres of the slits C$_{1}$ and C$_{2}$ and reduce the widths of the slits such that only  $C_{1}$
and $C_{2}$ are open. The phase differences
over the paths SC$_1$P through $\psi_{1}$ and  SC$_2$P through $\psi_{2}$ will then continuously vary along this sweep. 
To be undecidable and obey ternary logic, 
the total wave function would have to be completely
symmetrical with respect to $\zeta_{1}$ and $\zeta_{2}$, such that it would have to be
$\zeta_{1} + \zeta_{2}$. This is the reason why we have to do QM and calculate $|\zeta_{1} + \zeta_{2}|^{2}$. 
But simultaneously,  $\zeta_{1}$ and $\zeta_{2}$ would have to be mutually exclusive and satisfy binary logic 
because ``God would know''. All points were the phase difference is not $\pm{\pi\over{2}} ($mod $2\pi)$ should therefore have  zero amplitude
and not belong to the domain of $\zeta$ where $\zeta \neq 0$. This shows that  $\zeta$ must correspond to an interference pattern.
 From this point of view, an interference pattern (with its quantization of momentum ${\mathbf{p}} = h{\mathbf{q}}$)
 appears then as the only solution for the wave function that can satisfy 
simultaneously the requirements imposed by the binary and the ternary logic. 
The real experiment with non-zero slit widths
would not yield the Dirac comb but a blurred result due
to integration over the finite widths of the slits. A different point D$_{2} \neq$ C$_{2}$ might indeed provide an alternative path
SD$_2$P  that leads to the correct phase difference ${\pi\over{2}}$ with SC$_1$P. Let us call the width of the slits $w$ and  the wavelength $\lambda$.
The smaller the ratio $\lambda/w$ the easier it will be to find such points D$_{2}$. We can achieve this by increasing
 the energy of the electron, but this will eventually render the interactions incoherent such that we end up in the classical tennis ball regime.
 A far more interesting way is to change $\lambda/w$ by fiddling with the geometry of the set-up.
 Finally note that the phase difference of $\pm{\pi\over{2}} ($mod $2\pi)$
over the paths SS$_1$P through $\psi'_{1}$ and  SS$_2$P through $\psi'_{2}$ is not in contradiction with the phase difference $2\pi n$
over the paths SS$_1$P and  SS$_2$P through $\psi_{3}$ we discussed above, because $\psi_{3} = \psi'_{1} + \psi'_{2}$.

The true reason why we can calculate $\psi'_{3} = \psi'_{1} + \psi'_{2}$ as $\psi_{3} = \psi_{1} \boxplus \psi_{2}$ can within the Born approximation also be
explained by the linearity of the Fourier transform used in Eq. \ref{Born}  \cite{Coddens-HAL}, which is a better argument than invoking  the linearity of the wave equation. 
The path integral result is just a more refined equivalent of this, based on a more refined integral transform.
The reason for the presence of the Fourier transform in the formalism  is the fact that the electron spins \cite{Coddens-HAL,Coddens}. 
One can derive the whole wave formalism purely classically, just  from the assumption that the electron spins.
Eq. \ref{Born} hinges also crucially on the Born rule $p = |\psi|^{2}$. There is no rigorous proof for this rule
 but there exists ample justification for it \cite{Coddens-HAL}. 
The undecidability is completely due to the properties of the potential which defines both the local interactions and the global symmetry. 

The different fringes in the interference pattern are due to the fact that one of the electrons has traveled a longer path
than the other one such that it has made $n\in {\mathbb{N}}$ turns more. It is thus an ``older'' electron. This idea could be illustrated
by  making the experiment with muons in an experiment wherein the dimensions of the set-up are tuned with respect to the decay length.
 Due to the  time decay of the muons, the question through which one of the slits the muon has traveled
will become less undecidable in the wings of the distribution and
this will have an effect on the interference pattern. Conversely, one could imagine an interference experiment to measure a life time.

\section{Conclusion} \label{conclude}

In summary, we have proposed an intelligible solution for the paradox of the double-slit experiment.  
What we must learn from it is that there is no way one can use probabilities obtained  from one experiment
in the analysis of another experiment. The probabilities are conditional and context-bound.
Combining results from different contexts in a same calculation
should therefore be considered as taboo. In deriving Bell-type inequalities one transgresses
this taboo.
A much more detailed account of this work
is given in reference \cite{Coddens-HAL}, which fills many gaps and also explains how the same argument
of  context dependence can be applied to paradoxes related to Bell-type inequalities.

{\em Acknowledgments}. The author  thanks his director, Kees van de Beek for his continuous interest in this work and the CEA for the possibility to carry out this research.


\begin{thebibliography}{99}
\bibitem{note0}  One could also try to smooth out the  discontinuity mathematically, e.g. by taking
something in the spirit of $(1-e^{-|{\mathbf{K\cdot}} ({\mathbf{r}}-{\mathbf{r}}_{0})|})\,\psi_{3} + e^{-|{\mathbf{K\cdot}} ({\mathbf{r}}-{\mathbf{r}}_{0})|}\psi_{1} +  e^{-|{\mathbf{K\cdot}} ({\mathbf{r}}-{\mathbf{r}}_{0})|}\psi_{2}$, on lines ${\mathbf{r}} = {\mathbf{r}}_{0} + {\mathbf{K}} u, u \in {\mathbb{R}}$ through each point ${\mathbf{r}}_{0} \in \partial W$ of
 the boundary of $W$,  where $|{\mathbf{K}}|$ is very large
and where on each line, the vector ${\mathbf{K}}$ is taken perpendicular to the boundary of $W$.
The domain $W$ has at least two boundaries (there could be more due to the diffraction), but in each transition region only  one of the
 families of lines will give rise to a contribution $e^{-|{\mathbf{K\cdot}} ({\mathbf{r}}-{\mathbf{r}}_{0})|}$  that is not negligible. This is elaborate but
could be feasible because certain terms cancel one another. However, such a purely mathematical approach
is highly artificial and contains a lot of arbitrariness, which is difficult to remove on the basis
of  physical arguments. Accepting the discontinuity of the schematic description seems in this respect more physical.
\bibitem{note} This is a recurrent theme in QM \cite{Coddens-HAL,Coddens}.
Many conceptual problems of QM trace back to the poor understanding of the geometrical meaning of the spinors in the group theory.
The algebra is used as a black box, and in its philosophy to ``shut up and calculate'' it violates the meaning of the geometry while yielding
numerically correct results. But these geometrical errors may manifest themselves by
crystallizing under the form of some conceptual difficulties.
In binary logic, the presence of destructive interference $\psi_{1}({\mathbf{r}}) + \psi_{2}({\mathbf{r}}) = 0$ in the theory leads to insuperable conceptual difficulties,
which should have raised a red flag.
\end{thebibliography}
\end{document}